\documentclass[conference]{IEEEtran}

\IEEEoverridecommandlockouts

\usepackage{amsmath,amsthm,relsize}
\usepackage{amsfonts, amssymb, cuted}
\usepackage{cleveref}
\usepackage{mathtools}
\usepackage{algorithm}
\usepackage[noend]{algpseudocode}
\usepackage{cite}
\usepackage{xcolor}
\usepackage{soul}
\usepackage{graphicx}
\usepackage{tikz}
\usepackage{pgfplotstable}
\usepackage{pgfplots}
\usepackage{nicefrac}
\usepackage{enumitem}
\usepackage{support-caption}
\usepackage{subcaption}
\usepackage[font=footnotesize]{caption}
\usepackage{multirow}
\pgfplotsset{compat=1.15}
\usepackage{relsize}
\usetikzlibrary{patterns}
\usepackage{acro}

\newcommand{\vbar}{\raisebox{.17ex}{\rule{.04em}{1.35ex}}}
\newcommand{\vbarind}{\raisebox{.01ex}{\rule{.04em}{1.1ex}}}
\newcommand{\R}{\ifmmode{\rm I}\hspace{-.2em}{\rm R} \else ${\rm I}\hspace{-.2em}{\rm R}$ \fi}
\newcommand{\T}{\ifmmode{\rm I}\hspace{-.2em}{\rm T} \else ${\rm I}\hspace{-.2em}{\rm T}$ \fi}
\newcommand{\N}{\ifmmode{\rm I}\hspace{-.2em}{\rm N} \else \mbox{${\rm I}\hspace{-.2em}{\rm N}$} \fi}
\newcommand{\B}{\ifmmode{\rm I}\hspace{-.2em}{\rm B} \else \mbox{${\rm I}\hspace{-.2em}{\rm B}$} \fi}
\newcommand{\Hil}{\ifmmode{\rm I}\hspace{-.2em}{\rm H} \else \mbox{${\rm I}\hspace{-.2em}{\rm H}$} \fi}
\newcommand{\C}{\ifmmode\hspace{.2em}\vbar\hspace{-.31em}{\rm C} \else \mbox{$\hspace{.2em}\vbar\hspace{-.31em}{\rm C}$} \fi}
\newcommand{\Cind}{\ifmmode\hspace{.2em}\vbarind\hspace{-.25em}{\rm C} \else \mbox{$\hspace{.2em}\vbarind\hspace{-.25em}{\rm C}$} \fi}
\newcommand{\Q}{\ifmmode\hspace{.2em}\vbar\hspace{-.31em}{\rm Q} \else \mbox{$\hspace{.2em}\vbar\hspace{-.31em}{\rm Q}$} \fi}
\newcommand{\Z}{\ifmmode{\rm Z}\hspace{-.28em}{\rm Z} \else ${\rm Z}\hspace{-.28em}{\rm Z}$ \fi}

\DeclareAcronym{AWGN}{
    short = AWGN,
    long = additive white Gaussian noise,
    list = Additive White Gaussian Noise,
    tag = abbrev
}

\DeclareAcronym{ADMM}{
    short = ADMM,
    long = alternating direction method of multipliers,
    list = Alternating Direction Method of Multipliers,
    tag = abbrev
}

\DeclareAcronym{MGMC}{
    short = MGMC,
    long = multi-group multi-casting,
    list = multi-group multi-casting,
    tag = abbrev
}

\DeclareAcronym{SGMC}{
    short = SGMC,
    long = single-group multi-casting,
    list = single-group multi-casting,
    tag = abbrev
}
\DeclareAcronym{AoA}{
    short = AoA,
    long = angle-of-arrival,
    list = Angle-of-Arrival,
    tag = abbrev
}

\DeclareAcronym{AoD}{
    short = AoD,
    long = angle-of-departure,
    list = Angle-of-Departure,
    tag = abbrev
}

\DeclareAcronym{KKT}{
    short = KKT,
    long = Karush-Kuhn-Tucker,
    list = Karush-Kuhn-Tucker,
    tag = abbrev
}

\DeclareAcronym{MMF}{
    short = MMF,
    long = max-min-fairness,
    list = max-min-fairness,
    tag = abbrev
}

\DeclareAcronym{WMMF}{
    short = WMMF,
    long = weighted max-min-fairness,
    list = max-min-fairness,
    tag = abbrev
}

\DeclareAcronym{BB}{
    short = BB,
    long = base band,
    list = Base Band,
    tag = abbrev
}

\DeclareAcronym{BC}{
    short = BC,
    long = broadcast channel,
    list = Broadcast Channel,
    tag = abbrev
}

\DeclareAcronym{BS}{
    short = BS,
    long = base station,
    list = Base Station,
    tag = abbrev
}

\DeclareAcronym{BR}{
    short = BR,
    long = best response,
    list = Best Response, 
    tag = abbrev
}

\DeclareAcronym{CB}{
    short = CB,
    long = coordinated beamforming,
    list = Coordinated Beamforming,
    tag = abbrev
}

\DeclareAcronym{CC}{
    short = CC,
    long = coded caching,
    list = Coded Caching,
    tag = abbrev
}

\DeclareAcronym{CE}{
    short = CE,
    long = channel estimation,
    list = Channel Estimation,
    tag = abbrev
}

\DeclareAcronym{CoMP}{
    short = CoMP,
    long = coordinated multi-point transmission,
    list = Coordinated Multi-Point Transmission,
    tag = abbrev
}

\DeclareAcronym{CRAN}{
    short = C-RAN,
    long = cloud radio access network,
    list = Cloud Radio Access Network,
    tag = abbrev
}

\DeclareAcronym{CSE}{
    short = CSE,
    long = channel specific estimation,
    list = Channel Specific Estimation,
    tag = abbrev
}

\DeclareAcronym{CSI}{
    short = CSI,
    long = channel state information,
    list = Channel State Information,
    tag = abbrev
}

\DeclareAcronym{CSIT}{
    short = CSIT,
    long = channel state information at the transmitter,
    list = Channel State Information at the Transmitter,
    tag = abbrev
}

\DeclareAcronym{CU}{
    short = CU,
    long = central unit,
    list = Central Unit,
    tag = abbrev
}

\DeclareAcronym{D2D}{
    short = D2D,
    long = device-to-device,
    list = Device-to-Device,
    tag = abbrev
}

\DeclareAcronym{DE-ADMM}{
    short = DE-ADMM,
    long = direct estimation with alternating direction method of multipliers,
    list = Direct Estimation with Alternating Direction Method of Multipliers,
    tag = abbrev
}

\DeclareAcronym{DE-BR}{
    short = DE-BR,
    long = direct estimation with best response,
    list = Direct Estimation with Best Response,
    tag = abbrev
}

\DeclareAcronym{DE-SG}{
    short = DE-SG,
    long = direct estimation with stochastic gradient,
    list = Direct Estimation with Stochastic Gradient,
    tag = abbrev
}

\DeclareAcronym{DFT}{
	short = DFT,
	long = discrete fourier transform,
	list = Discrete Fourier Transform,
	tag = abbrev
}

\DeclareAcronym{DoF}{
    short = DoF,
    long = degrees of freedom,
    list = Degrees of Freedom,
    tag = abbrev
}

\DeclareAcronym{DL}{
    short = DL,
    long = downlink,
    list = Downlink,
    tag = abbrev
}

\DeclareAcronym{GD}{
	short = GD, 
	long = gradient descent,
	list = Gradeitn Descent,
	tag = abbrev
}

\DeclareAcronym{IBC}{
    short = IBC,
    long = interfering broadcast channel,
    list = Interfering Broadcast Channel,
    tag = abbrev
}

\DeclareAcronym{i.i.d.}{
    short = i.i.d.,
    long = independent and identically distributed,
    list = Independent and Identically Distributed,
    tag = abbrev
}

\DeclareAcronym{JP}{
    short = JP,
    long = joint processing,
    list = Joint Processing,
    tag = abbrev
}

\DeclareAcronym{LOS}{
	short = LOS,
	long = line-of-sight,
	list = Line-of-Sight,
	tag = abbrev
}

\DeclareAcronym{LS}{
    short = LS,
    long = least squares,
    list = Least Squares,
    tag = abbrev
}

\DeclareAcronym{LTE}{
    short = LTE,
    long = Long Term Evolution,
    tag = abbrev
}

\DeclareAcronym{LTE-A}{
    short = LTE-A,
    long = Long Term Evolution Advanced,
    tag = abbrev
}

\DeclareAcronym{MIMO}{
    short = MIMO,
    long = multiple-input multiple-output,
    list = Multiple-Input Multiple-Output,
    tag = abbrev
}

\DeclareAcronym{MISO}{
    short = MISO,
    long = multiple-input single-output,
    list = Multiple-Input Single-Output,
    tag = abbrev
}

\DeclareAcronym{MAC}{
    short = MAC,
    long = multiple access channel,
    list = Multiple Access Channel,
    tag = abbrev
}

\DeclareAcronym{MSE}{
    short = MSE,
    long = mean-squared error,
    list = Mean-Squared Error,
    tag = abbrev
}

\DeclareAcronym{MMSE}{
    short = MMSE,
    long = minimum mean-squared error,
    list = Minimum Mean-Squared Error,
    tag = abbrev
}

\DeclareAcronym{mmWave}{
	short = mmWave,
	long = millimeter wave,
	list = Millimeter Wave,
	tag = abbrev
}

\DeclareAcronym{MU-MIMO}{
    short = MU-MIMO,
    long = multi-user \ac{MIMO},
    list = Multi-User \ac{MIMO},
    tag = abbrev
}

\DeclareAcronym{OTA}{
    short = OTA,
    long = over-the-air,
    list = Over-the-Air,
    tag = abbrev
}

\DeclareAcronym{PSD}{
    short = PSD,
    long = positive semidefinite,
    list = Positive Semidefinite,
    tag = abbrev
}

\DeclareAcronym{QoS}{
	short = QoS,
	long = quality of service,
	list = Quality of Service,
	tag = abbrev
}

\DeclareAcronym{RCP}{
	short = RCP,
	long = remote central processor,
	list = Remote Central Processor,
	tag = abbrev
}

\DeclareAcronym{RRH}{
    short = RRH,
    long = remote radio head,
    list = Remote Radio Head,
    tag = abbrev
}

\DeclareAcronym{RSSI}{
    short = RSSI,
    long = received signal strength indicator,
    list = Received Signal Strength Indicator,
    tag = abbrev
}

\DeclareAcronym{RX}{
	short = RX,
	long = receiver,
	list = Receiver,
	tag = abbrev
}

\DeclareAcronym{SCA}{
    short = SCA,
    long = successive-convex-approximation,
    list = Successive-Convex-Approximation,
    tag = abbrev
}

\DeclareAcronym{SG}{
    short = SG,
    long = stochastic gradient,
    list = Stochastic Gradient,
    tag = abbrev
}

\DeclareAcronym{SIC}{
    short = SIC,
    long = successive interference cancellation,
    list = Successive Interference Cancellation,
    tag = abbrev
}

\DeclareAcronym{SNR}{
    short = SNR,
    long = signal-to-noise-ratio,
    list = Signal-to-Noise Ratio,
    tag = abbrev
}

\DeclareAcronym{SDR}{
    short = SDR,
    long = semi-definite-relaxation,
    list = semi-definite-relaxation,
    tag = abbrev
}

\DeclareAcronym{SINR}{
    short = SINR,
    long = signal-to-interference-plus-noise ratio,
    list = Signal-to-Interference-plus-Noise Ratio,
    tag = abbrev
}

\DeclareAcronym{SOCP}{
	short = SOCP, 
	long = second order cone program,
	list = Second Order Cone Program,
	tag = abbrev
}

\DeclareAcronym{SSE}{
    short = SSE,
    long = stream specific estimation,
    list = Stream Specific Estimation,
    tag = abbrev
}

\DeclareAcronym{SVD}{
	short = SVD,
	long = singular value decomposition,
	list = Singular Value Decomposition,
	tag = abbrev
}

\DeclareAcronym{TDD}{
	short = TDD,
	long = time division duplex,
	list = Time Division Duplex,
	tag = abbrev
}

\DeclareAcronym{TX}{
	short = TX,
	long = transmitter,
	list = Transmitter,
	tag = abbrev
}

\DeclareAcronym{UE}{
    short = UE,
    long = user equipment,
    list = User Equipment,
    tag = abbrev
}

\DeclareAcronym{UL}{
    short = UL,
    long = uplink,
    list = Uplink,
    tag = abbrev
}

\DeclareAcronym{ULA}{
	short = ULA,
	long = uniform linear array,
	list = Uniform Linear Array,
	tag = abbrev
}

\DeclareAcronym{UPA}{
    short = UPA,
    long = uniform planar array,
    list = Uniform Planar Array,
    tag = abbrev
}

\DeclareAcronym{WMMSE}{
    short = WMMSE,
    long = weighted minimum mean-squared error,
    list = Weighted Minimum Mean-Squared Error,
    tag = abbrev
}

\DeclareAcronym{WMSEMin}{
    short = WMSEMin,
    long = weighted sum \ac{MSE} minimization,
    list = Weighted sum \ac{MSE} Minimization,
    tag = abbrev
}

\DeclareAcronym{WBAN}{
	short = WBAN,
	long = wireless body area network,
	list = Wireless Body Area Network,
	tag = abbrev
}

\DeclareAcronym{WSRMax}{
    short = WSRMax,
    long = weighted sum rate maximization,
    list = Weighted Sum Rate Maximization,
    tag = abbrev
}

\theoremstyle{definition}

\newtheorem{exmp}{Example}

\newcommand{\CF}[0]{{\mathcal{F}}}

\newcommand{\CR}[0]{{\mathcal{R}}}
\newcommand{\CS}[0]{{\mathcal{S}}}
\newcommand{\CT}[0]{{\mathcal{T}}}

\newcommand{\Bh}[0]{{\mathbf{h}}}

\newcommand{\Bu}[0]{{\mathbf{u}}}
\newcommand{\Bv}[0]{{\mathbf{v}}}
\newcommand{\Bw}[0]{{\mathbf{w}}}
\newcommand{\Bx}[0]{{\mathbf{x}}}
\newcommand{\By}[0]{{\mathbf{y}}}
\newcommand{\Bz}[0]{{\mathbf{z}}}

\newcommand{\BH}[0]{{\mathbf{H}}}
\newcommand{\BI}[0]{{\mathbf{I}}}

\newcommand{\Sfs}[0]{{\mathsf{s}}}

\newcommand{\SfG}[0]{{\mathsf{G}}}

\newcommand{\SfL}[0]{{\mathsf{L}}}

\newcommand{\FillGray}[3]{\filldraw[gray!50](#3-1+0.1,#1-#2+0.1) rectangle (#3-0.1,#1-#2+1-0.1)}
\newcommand{\FillBlack}[3]{\filldraw[black!70](#3-1+0.1,#1-#2+0.1) rectangle (#3-0.1,#1-#2+1-0.1)}

\newcommand{\PutText}[4]{\node[] at (#3-1+0.5,#1-#2+0.5) {\small #4}}

\newcommand{\subparagraph}{}
\usepackage{titlesec}
\titlespacing\section{3pt}{6pt plus 4pt minus 2pt}{6pt plus 2pt minus 2pt}
\titlespacing\subsection{3pt}{4pt plus 4pt minus 2pt}{4pt plus 2pt minus 2pt}
\titlespacing\subsubsection{3pt}{3pt plus 4pt minus 2pt}{0pt plus 2pt minus 3pt}

\title{A Low-Subpacketization High-Performance\\ MIMO Coded Caching Scheme 
}

\begin{document}

\author{\IEEEauthorblockN{MohammadJavad Salehi, Hamidreza Bakhshzad Mahmoodi and Antti T\"olli} \\
\IEEEauthorblockA{
    Centre for Wireless Communications, University of Oulu, 90570 Oulu, Finland \\
    \textrm{E-mail: \{firstname.lastname\}@oulu.fi}
    }
\thanks{
This work is supported by the Academy of Finland under grants no. 319059 (Coded Collaborative Caching for Wireless Energy Efficiency) and 318927 (6Genesis Flagship), and by Vaikuttavuuss\"a\"ati\"o under the project Directional Data Delivery for Wireless Immersive Digital Environments (3D-WIDE).}
}

\maketitle

\begin{abstract}
In this paper, we study how coded caching can be efficiently applied to multiple-input multiple-output (MIMO) communications. This is an extension to cache-aided multiple-input single-output (MISO) communications, where it is shown that with an $L$-antenna transmitter and coded caching gain $t$, a cumulative coded caching and spatial multiplexing gain of $t+L$ is achievable. We show that, interestingly, for MIMO setups with $G$-antenna receivers, a coded caching gain larger than MISO setups by a multiplicative factor of $G$ is possible, and the full coded caching and spatial multiplexing gain of $Gt+L$ is also achievable. Furthermore, we propose a novel algorithm for building low-subpacketization, high-performance MIMO coded caching schemes using a large class of existing MISO schemes.

\end{abstract}

\begin{IEEEkeywords}
coded caching, MIMO communications, low-subpacketization
\end{IEEEkeywords}

\section{Introduction}
Wireless communication networks are under mounting pressure to support exponentially increasing volumes of multimedia content~\cite{cisco2018cisco}, as well as to support the imminent emergence of new applications such as wireless immersive viewing~\cite{mahmoodi2021non}. 
For the efficient delivery of such multimedia content, the work of  Maddah-Ali and Niesen in~\cite{maddah2014fundamental} proposed the idea of \emph{coded caching} as a means of increasing the data rates by exploiting cache content across the network.
In a single-stream downlink network of $K$ cache-enabled users who can pre-fetch data into their cache memories, coded caching enables boosting the achievable rate by a multiplicative factor proportional to the cumulative cache size in the entire network.
The key to achieving this speedup lies in the multicasting of carefully created codewords to different groups of users, such that each user can use its cache content to remove unwanted parts (i.e., the interference) from the received signal.
With a library of $N$ equal-sized files, and given that each user can pre-fetch $M$ files into its cache memory during the so-called \textit{placement} phase, coded caching enables serving groups of users of size $t+1$ with any single transmission, where $t:=\frac{KM}{N}$ is called the \textit{coded caching gain}.
In other words, with coded caching one can include $t+1$ independent data streams within a single transmission, and hence, the achievable \ac{DoF} is increased from one to $t+1$.

Motivated by the growing importance of multi-antenna wireless communications~\cite{rajatheva2020white}, Shariatpanahi et al.~\cite{shariatpanahi2016multi,shariatpanahi2018physical} explored the cache-aided \ac{MISO} setting, for which it revealed that the same coded caching gain $t$ could be achieved cumulatively with the spatial multiplexing gain. More exactly, for a downlink \ac{MISO} setup with the transmitter-side multiplexing gain of $L$, the work in~\cite{shariatpanahi2018physical} developed a method that achieved a \ac{DoF} of $t+L$, which was later shown in~\cite{lampiris2018resolving} to be optimal under basic assumptions of uncoded cache placement and one-shot linear data delivery. Of course, an even larger cache-aided \ac{DoF} can be achieved for the same setup using interference-alignment techniques such as in~\cite{naderializadeh2017fundamental}. However, such techniques are overly complex to implement in practice (cf.~\cite{maddah2014fundamental}), and hence, in this paper, we also aim at maximizing the \ac{DoF} with the underlying assumptions of uncoded cache placement and one-shot linear data delivery.

Following the introduction of single- and multi-stream coded caching in~\cite{maddah2014fundamental,shariatpanahi2018multi}, many later works in the literature addressed their important scaling and performance issues. For \ac{MISO} setups, in~\cite{tolli2017multi}, the authors showed that using optimized beamformers instead of zero-forcing, one could achieve a better rate for communications at the finite \ac{SNR} regime. Also in the same paper, two interesting methods for adjusting the \ac{DoF} and controlling the \ac{MAC} size were introduced for managing the optimized beamformer design complexity. 
Similarly, the exponentially growing subpacketization issue (i.e., the number of smaller parts each file should be split into) was addressed thoroughly in the literature, for both single-stream (cf.~\cite{yan2018placement,yan2017placement,shangguan2018centralized}) and multi-stream (cf.\cite{lampiris2018adding,salehi2020lowcomplexity,salehi2019coded,salehi2020diagonal,mohajer2020miso}) setups. Most notably, for multi-stream \ac{MISO} setups it was shown in~\cite{lampiris2018adding,salehi2019coded,salehi2020lowcomplexity} that using \textit{signal-level} interference cancellation instead of the \textit{bit-level} approach, one could substantially reduce the subpacketization requirement while achieving the same maximum \ac{DoF} value of $t+L$. The difference between signal- and bit-level approaches is that with the former, cache-aided interference cancellation is done prior to decoding, while with the latter, it is performed after the received signal is decoded. Of course, as shown in~\cite{salehi2019subpacketization,salehi2020coded}, signal-level interference cancellation modestly decreases the achievable rate (with respect to bit-level approaches), especially at the finite-\ac{SNR} regime. Nevertheless, it does not harm the achievable \ac{DoF} and yet enables simpler optimized beamformer designs, as discussed in~\cite{salehi2020lowcomplexity}.

In this paper, we extend the applicability of coded caching to \ac{MIMO} communications with multi-stream receivers. We show that \ac{MIMO} coded caching can be considered as an extension of the shared-cache setting studied in~\cite{parrinello2019fundamental,parrinello2020extending}. In shared-cache setups, a number of single-stream users are aided by a smaller number of shared-cache nodes (which can be, e.g., helper nodes) while requesting data from a multi-stream transmitting server. 
To extend this model to \ac{MIMO} coded caching, we keep the transmitter in place and replace each shared-cache node with a multi-stream receiver. Then, if the number of attached users to the shared-cache node $k$ is $G_k$, we set the receiver-side multiplexing gain of its equivalent multi-stream receiver to be $G_k$.
%
Such an extension paves the way for using the optimality results in~\cite{parrinello2019fundamental,parrinello2020extending} to \ac{MIMO} coded caching. Specifically, in case all the users in a \ac{MIMO} setup can receive $G$ independent streams (which is the case, for example, when they have $G$ adequately-separated antennas) and $\eta := \frac{L}{G}$ is an integer, we get the interesting result that the coded caching gain $t$ can be multiplied by $G$, and a cumulative single-shot \ac{DoF} of $Gt+L$ becomes achievable.
We also propose a new algorithm for creating \ac{MIMO} coded caching schemes, where we elevate baseline linear \ac{MISO} schemes already available in the literature (with the signal-level cache cancellation approach) to be applicable in \ac{MIMO} setups while achieving the increased \ac{DoF} of $Gt+L$. This enables designing low-complexity, low-subpacketization \ac{MIMO} schemes with full \ac{DoF} using appropriate baseline \ac{MISO} schemes such as the one in~\cite{salehi2020lowcomplexity}.

In this paper, we use the following notations. For any integer $J$, we use $[J]$ to denote the set of numbers $\{1,2,\cdot \cdot \cdot,J\}$. 
Boldface upper- and lower-case letters indicate matrices and vectors, respectively, and
calligraphic letters denote sets. 
Other notations are defined as they are used throughout the text.

\section{System Model}
\label{section:sys_model}
We consider a MIMO communication setup with $K$ users and a single server, where the server has $\SfL$ transmit antennas and every user has $\SfG$ receive antennas. 
These antenna arrays enable spatial multiplexing gains of $L$ and $G$ to be achievable at the transmitter and receiver side, respectively. In other words, if the $\SfG \times \SfL$ channel matrix from the transmitter to user $k \in [K]$ is shown by $\BH_k$, the rank of every $\BH_k$ is at least $G$, and the rank of the cumulative channel matrix, defined by the vertical concatenation of individual channel matrices as
\begin{equation}
\BH =
\begin{bmatrix}
    \BH_1 \\
    \cdot \cdot \cdot \\
    \BH_K
\end{bmatrix}
\; ,
\end{equation}
is at least $L$. As a result, the server can deliver at least $L$ independent data streams to all the users, and each user can receive at least $G$ independent streams. Let us denote the data stream $g \in [G]$ at user $k \in [K]$ with $\Sfs_k^g$. Then,  transmit beamformers $\Bw_{\CR} \in \mathbb{C}^{\SfL \times 1}$ and receive beamformers $\Bu_{k,g} \in \mathbb{C}^{\SfG \times 1}$ can be built such that the zero-forcing set $\CR$ includes $L-1$ streams, and for every $\Sfs_{k}^{g} \in \CR$ we have
\begin{equation}
    \Bu_{k,g}^H \BH_{k} \Bw_{\CR} = 0 \; .
\end{equation}
As the size of $\CR$ is $L-1$, each transmit beamformer removes the interference on $L-1$ streams. 
For simplicity, we define 
\begin{equation}
\label{eq:equiv_channel}
\Bh_{k,g} = (\Bu_{k,g}^H \BH_{k})^H 
\end{equation}
as the equivalent channel vector for the stream $\Sfs_k^g$. 
In this paper, we assume that equivalent channel multipliers (i.e., $\Bh_{k,g}^H \Bw_{\CR}$) can be estimated at the receivers.\footnote{This can be done, for example, using downlink precoded pilots.}

The users request files from a library $\CF$ of $N$ files, each with size $F$ bits. Each user has a cache memory of size $MF$ bits. For notational simplicity, we use a normalized data unit and ignore $F$ in subsequent notations. The coded caching gain is defined as $t = \frac{KM}{N}$. The value of $t$ also indicates how many copies of the file library can be stored in the cache memories throughout the entire network. 

The system operation consists of two phases, placement and delivery. During the placement phase, which takes place at the network's low-traffic time, cache memories of the users are filled with data from the files in $\CF$. The placement phase is done without any prior knowledge of the file requests in the future, and hence, we assume equal-sized portions of all files are stored in each cache memory. Then, during the delivery phase, the users reveal their requested files and the server builds and transmits a set of transmission vectors so that all the users can reconstruct their requested files using the received data and their cached contents. Without loss of generality, we assume the transmissions are done in a TDMA manner, i.e., in consecutive time intervals.

In this paper, we focus on high-\ac{SNR} communications and use \ac{DoF} in terms of the number of simultaneously delivered independent data streams as the performance metric.\footnote{Note that the common practice in the literature for MISO setups is to define DoF as the number of simultaneously served \textit{users} instead of \textit{streams}, as in a MISO setup each user can receive only one stream at a time.} We also assume that $\eta = \frac{L}{G}$ is an integer. Let us denote the transmission vector for time interval $i$ with $\Bx(i)$ and the set of transmitted streams in this time interval as $\CS(i)$. The vector $\Bx(i)$ is built as
\begin{equation}
    \Bx(i) = \sum_{\Sfs_k^g \in \CS(i)} X_k^g(i) \Bw_{\CR_k^g(i)} \; ,
\end{equation}
where $X_k^g(i)$ is the data content for the stream $\Sfs_k^g$ at time interval $i$ and $\CR_k^g(i)$ denotes its associated zero-forcing set. After the transmission is concluded, the user $k$ receives
\begin{equation}
    \By_k(i) = \sum_{\Sfs_k^g \in \CS(i)} X_k^g(i) \BH_k \Bw_{\CR_k^g(i)} + \Bz_k(i) \; ,
\end{equation}
where the vector $\Bz_k(i) \in \mathbb{C}^{\SfG \times 1}$ represents \ac{AWGN} at user $k$ in time interval $i$. Then, after multiplying by receive beamformer vectors $\Bu_{k,g}$, the data stream $g$ at user $k$ in this time interval can be found by decoding the signal
\begin{equation}
    \begin{aligned}
    y_k^g(i) &= \sum_{\Sfs_k^g \in \CS(i)} X_k^g(i) \Bu_{k,g}^H \BH_k \Bw_{\CR_k^g(i)} + z_{k,g}(i) \\
    & =  \sum_{\Sfs_k^g \in \CS(i)} X_k^g(i) \Bh_{k,g}^H \Bw_{\CR_k^g(i)} + z_{k,g}(i) \; ,
    \end{aligned}
\end{equation}
where we have used $z_{k,g}(i) \equiv \Bu_{k,g}^H \Bz_k(i)$. The goal is to design cache placement and transmission vectors such that for every $\Sfs_k^g \in \CS(i)$, $X_k^g(i)$ can be decoded interference-free from $y_k^g(i)$. As mentioned earlier, the performance metric is the number of simultaneously delivered independent data streams within each transmission, and hence, we want to maximize $|\CS(i)|$ for every time interval. The maximum value of $|\CS(i)|$ for a \ac{MISO} setup is shown to be $t+L$. In this paper, we show that an increased \ac{DoF} value of $Gt+L$ is achievable in a \ac{MIMO} setup with a low subpacketization complexity.

\section{Signal-level Interference Cancellation}
The \ac{MISO} coded caching scheme in~\cite{lampiris2018adding} introduced a novel cache-aided interference cancellation approach, different than~\cite{shariatpanahi2018physical}, to reduce the subpacketization without altering the maximum achievable \ac{DoF}. This approach, which we refer to as the signal-level approach, removes the interference using the cache contents \textit{before} the signal is decoded at the receiver. This is in contrast with the classical bit-level approach, where the cache-aided interference removal is done \textit{after} decoding the received signal. Many other papers in the literature, including the one in~\cite{salehi2020lowcomplexity}, have also used the signal-level approach to reduce the required subpacketization. It is worth mentioning that, although we consider only the subpacketization issue here, the signal-level approach also has other benefits such as enabling much simpler optimized beamformer designs, as demonstrated in~\cite{salehi2020lowcomplexity}.

The \ac{MIMO} coded caching scheme proposed in this paper can be built using any baseline \ac{MISO} scheme with the signal-level interference cancellation approach. Interestingly though, with minor modifications, any bit-level \ac{MISO} coded caching scheme can be transformed into an equivalent signal-level \ac{MISO} scheme, without any DoF loss. The key to this transformation is to decouple the codewords and use a separate beamformer for every data element being sent. The following example enlightens the difference between the signal- and bit-level approaches, and shows how a bit-level scheme can be transformed to an equivalent signal-level scheme.

\begin{exmp}
\label{exmp:sig_bit_level}
Consider a small {MISO} setup with $K=3$ users where the coded caching gain is $t=1$ and the spatial multiplexing gain of $L=2$ can be achieved at the transmitter. Let us consider the bit-level coded caching scheme in~\cite{shariatpanahi2018physical} as the baseline. This scheme requires each file $W \in \CF$ to be split into three subpackets $W_p$, $p \in [3]$. The user $k$, $k \in [3]$, stores subpackets $W_k$ of every file $W$ in its cache memory; i.e., the cache contents of user $k$ are $\{W_k;\;\forall W\in\CF\}$. Then, during the delivery phase, if the files requested by users 1-3 are shown by $A$, $B$, and $C$, respectively, we transmit
\begin{equation}
    \Bx = (A_2 \oplus B_1) \Bw_3 + (A_3 \oplus C_1) \Bw_2 + (B_3 \oplus C_2) \Bw_1 \; ,
\end{equation}
where $\oplus$ represents the bit-wise XOR operation, and $\Bw_k$ is the zero-forcing vector for user $k$ (i.e., $\Bh_k^H \Bw_k = 0$, where $\Bh_k$ is the channel vector from the transmitter to user $k$). Without loss of generality, let us consider the decoding process at user one. Following the zero-force beamforming definition, this user receives
\begin{equation}
    y_1 = (A_2 \oplus B_1) \Bh_1^H \Bw_3 + (A_3 \oplus C_1) \Bh_1^H \Bw_2 + z_1 \; ,
\end{equation}
where $z_1$ represents the \ac{AWGN} at user one. The next step is to decode both $(A_2 \oplus B_1)$ and $(A_3 \oplus C_1)$ from $y_1$, using, for example, a \ac{SIC} receiver. Finally, \emph{after} decoding, user one has to use its cache contents to remove $B_1$ and $C_1$ from the resulting terms, to get $A_2$ and $A_3$ interference-free. As the cache-aided interference cancellation is done after the received signal is decoded at the receiver, we have a bit-level coded caching scheme here. 

In order to transform this scheme into an equivalent signal-level scheme, we need to simply decouple the codewords (i.e., the XOR terms) and use a separate beamformer for every single data element. This means, instead of the transmission vector $\Bx$, we transmit
\begin{equation}
    \Bar{\Bx} = A_2 \Bv_3^1 + B_1 \Bw_3^2 + A_3 \Bw_2^1 + C_1 \Bw_2^2 + B_3 \Bw_1^1 + C_2 \Bw_1^2 \; ,
\end{equation}
where the superscripts are used to distinguish two beamformers with the same zero-forcing set (but different powers). After the transmission of $\Bar{\Bx}$, user one receives
\begin{equation}
    \Bar{y}_1 =  A_2 \Bh_1^T \Bw_3^1 + B_1 \Bh_1^T \Bw_3^2 + A_3 \Bh_1^T \Bw_2^1 + C_1 \Bh_1^T \Bw_2^2 + z_1 \; .
\end{equation}
Then, in order to get its required data elements, i.e., $A_2$ and $A_3$, user one has to first build $B_1 \Bh_1^H \Bw_3^2$ and $C_1 \Bh_1^H \Bw_2^2$ using its cache contents and the acquired channel multiplier information, and remove these interference terms from the received signal \emph{before} decoding it (otherwise, these interference terms will degrade the achievable rate severely). As the cache-aided interference cancellation is now done before the decoding process, this second scheme is using a signal-level approach. 
\end{exmp}

\begin{figure}
    \begin{subfigure}{\columnwidth}
        \centering
        \includegraphics[width = \textwidth]{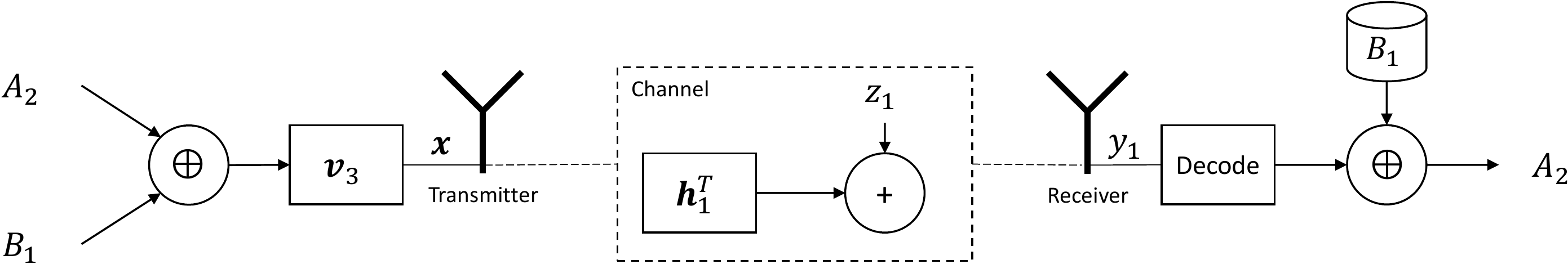}
        \caption{Bit-Level Interference Cancellation}
        \label{fig:subfig_bitlevel}
    \end{subfigure}
    \begin{subfigure}{\columnwidth}
        \centering
        \includegraphics[width = \textwidth]{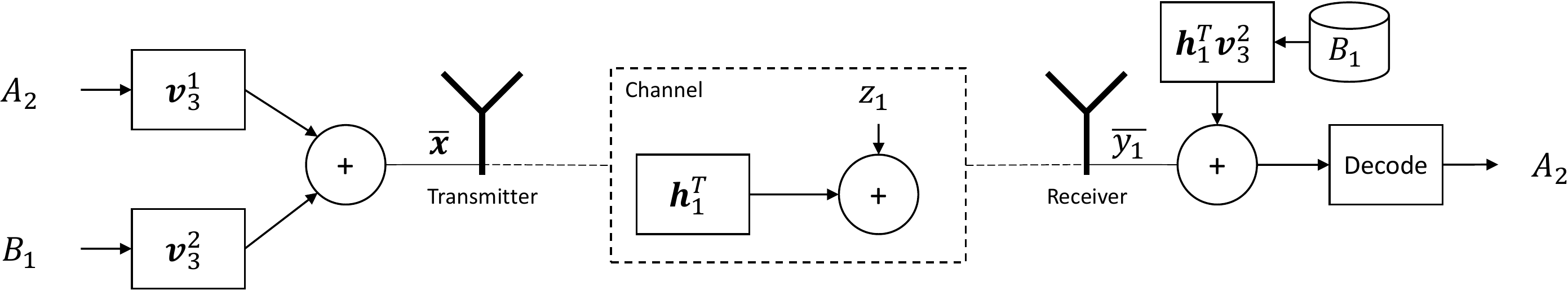}
        \caption{Signal-Level Interference Cancellation}
        \label{fig:subfig_siglevel}
    \end{subfigure}
    \caption{Bit- and Signal-Level Schemes in Example~\ref{exmp:sig_bit_level}}
    \label{fig:bit_sig_level}
\end{figure}

For better clarification, the encoding and decoding processes for the proposed bit- and signal-level schemes in Example~\ref{exmp:sig_bit_level} are shown in Figure~\ref{fig:bit_sig_level}. Note that in this figure, the decoding process for both schemes is shown only for user one. With the same procedure outlined in Example~\ref{exmp:sig_bit_level}, we can transform any \ac{MISO} scheme with the bit-level approach to another one with the signal-level approach. Hence, although our \ac{MIMO} coded caching scheme requires a baseline signal-level \ac{MISO} scheme, this condition does not in fact limit its applicability. Nevertheless, throughout the rest of this paper, we use the scheme in~\cite{salehi2020lowcomplexity}, due to its lower subpacketization, as our baseline scheme.

\section{MIMO Coded Caching}
\subsection{The Elevation Procedure}
\label{section:stretching_algorithm}
As stated earlier, our new algorithm for \ac{MIMO} coded caching works through elevation of a baseline signal-level \ac{MISO} scheme.
In a nutshell, this elevation procedure works as follows: for a given $K$-user \ac{MIMO} network with coded caching gain $t$ and spatial multiplexing gains of $L$ and $G$ at the transmitter and receivers, 
we first consider a virtual \ac{MISO} network with the same number of $K$ users and caching gain $t$, but with the spatial multiplexing gain of $\eta = \frac{L}{G}$ at the transmitter. Then, we design cache placement and transmission vectors for the virtual network following one of the available signal-level \ac{MISO} schemes in the literature. Finally, cache contents for users in the real network are set to be the same as the virtual network, but transmission vectors of the virtual network are \textit{stretched} to support multi-stream receivers and the increased \ac{DoF}. 

The stretching process of the transmission vectors of the virtual network is straightforward. Assume the zero-forcing beamformers for the virtual network are shown with $\Bw_{\CT}$, where $\CT$ denotes the set of users where the data is zero-forced.\footnote{Note that the virtual network has a \ac{MISO} setup, and hence, each user can receive only one data stream at a time. Moreover, virtual beamformer vectors $\Bw_{\CT}$ are defined only to clarify the transmission (and decoding) process in the virtual network, and hence, it is not necessary to calculate their exact values at any step.} Then, a transmission vector $\hat{\Bx}$ of the virtual network is a linear combination of $t+\eta$ terms $W_p^q(k) \Bw_{\CT(k)}$, where $W_p^q(k)$ denotes a subpacket of the file $W(k)$ requested by user $k$ (the subpacket index is clarified by $p$ and $q$), and $\Bw_{\CT(k)}$ is the beamformer vector associated with $W_p^q(k)$. Now, for stretching, for every term $W_p^q(k) \Bw_{\CT(k)}$ in $\hat{\Bx}$, we
%
%
first split the subpacket $W_p^q(k)$ into $G$ smaller parts $W_p^{q,g} (k)$, $g \in [G]$, and then replace the term with
\begin{equation}
\label{eq:stretch_main}
    W_p^q(k) \Bw_{\CT(k)} \; \rightarrow \; W_p^{q,1} (k) \Bw_{\CR_1} +  \cdot \cdot \cdot + W_p^{q,G} (k) \Bw_{\CR_G} ,
\end{equation}
where the zero-forcing sets $\CR_g$ for the real \ac{MIMO} network are built as
\begin{equation}
\label{eq:stretch_zfset}
    \CR_g = \bigcup_{\Bar{k} \in \CT(k)} \{ \Sfs_{\Bar{k}}^1, \cdot \cdot \cdot , \Sfs_{\Bar{k}}^G \} \; \cup \; \{ \Sfs_k^1, \cdot \cdot \cdot , \Sfs_k^G \} \backslash \{ \Sfs_k^g \} .
\end{equation}
In other words, instead of a single data stream $W_p^q(k)$ for user $k$ in the virtual network, we consider $G$ parallel streams $W_p^{q,g}(k)$ for user $k$ in the real network. Moreover, if the zero-forcing set for $W_p^q (k)$ in the virtual network is $\CT(k)$, the interference from $W_p^{q,g}(k)$ in the real network should be zero-forced at:
\begin{enumerate}
    \item Every stream sent to every user $\Bar{k} \in \CT(k)$. As $|\CT(k)| = \eta-1$ and each stream in the virtual network is transformed into $G$ streams in the real network, this includes $G \times (\eta-1)$ streams in total;
    \item Every other stream sent to user $k$. This includes $G-1$ streams in total.
\end{enumerate}
As a result, the interference from $W_p^{q,g}(k)$ must be zero-forced on a total number of
\begin{equation}
    G (\eta-1) + G-1 = \eta G -1 = L-1 
\end{equation}
streams in the real network, which is indeed possible as the spatial multiplexing gain of $L$ is achievable at the transmitter. In the following subsection, we review the elevation procedure using a simple example network. We also discuss how the received signal is decoded by the users.

\subsection{A Clarifying Example}
Let us consider a small network with $K=6$ users and caching gain $t=1$, where the spatial multiplexing gains of $L=4$ and $G=2$ are achievable at the transmitter and receivers, respectively.
Using \ac{MISO} schemes, the maximum achievable \ac{DoF} for this network is $L+t = 5$.\footnote{Note that in this case, multiple receive antennas can be used for achieving an additional beamforming gain (cf.~\cite{tolli2017multi}). However, they do not result in any increase in the \ac{DoF} value.} However, with the proposed elevation algorithm, the achievable \ac{DoF} can be increased to $Gt+L=6$. As mentioned earlier, without any loss of generality, we assume the reduced-subpacketization scheme in~\cite{salehi2020lowcomplexity} is chosen as the baseline \ac{MISO} scheme.

The first step is to find proper parameters for the virtual \ac{MISO} network. Clearly, the virtual network has the same number of $K=6$ users and caching gain $t=1$, but the spatial multiplexing gain of $\eta = \frac{L}{G} = 2$ at the transmitter. Cache placement in the virtual network directly follows that of~\cite{salehi2020lowcomplexity}, i.e., every file $W \in \CF$ is first split into $K=6$ equal-sized \textit{packets} denoted by $W_p$, $p \in [6]$, and then, each packet is split into $t+\eta=3$ smaller \textit{subpackets} denoted by $W_p^q$, $q \in [3]$. Finally, every user $k$ caches $W_k^q$, for every $W \in \CF$ and $q \in [3]$. For example, user one caches $\{W_1^1, W_1^2, W_1^3 \}$, for every file $W$ in the library.

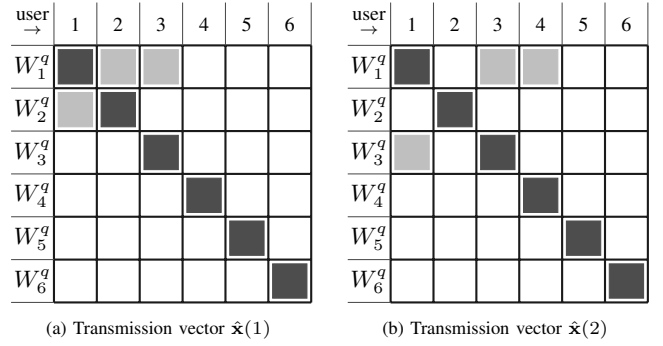
\begin{figure}[t]
    \centering
    \begin{subfigure}{.24\textwidth}
        \centering
        \begin{tikzpicture}[scale = 0.57]
            \begin{scope}<+->;
            \draw[step=1cm,thick,black!90] (1,0) grid (7,6);
            \draw[thin,black!70](0,0)to(1,0);
            \draw[thin,black!70](0,1)to(1,1);
            \draw[thin,black!70](0,2)to(1,2);
            \draw[thin,black!70](0,3)to(1,3);
            \draw[thin,black!70](0,4)to(1,4);
            \draw[thin,black!70](0,5)to(1,5);
            \draw[thin,black!70](0,6)to(1,6);
            \draw[thin,black!70](1,6)to(1,7);
            \draw[thin,black!70](2,6)to(2,7);
            \draw[thin,black!70](3,6)to(3,7);
            \draw[thin,black!70](4,6)to(4,7);
            \draw[thin,black!70](5,6)to(5,7);
            \draw[thin,black!70](6,6)to(6,7);
            \draw[thin,black!70](7,6)to(7,7);
            \end{scope}
            \begin{scope}

            \FillBlack{7}{2}{2};
            \FillBlack{7}{3}{3};
            \FillBlack{7}{4}{4};
            \FillBlack{7}{5}{5};
            \FillBlack{7}{6}{6};
            \FillBlack{7}{7}{7};
            
            \FillGray{7}{2}{3};
            \FillGray{7}{2}{4};
            \FillGray{7}{3}{2};
            
            \PutText{7}{2}{1}{$W_1^q$};
            \PutText{7}{3}{1}{$W_2^q$};
            \PutText{7}{4}{1}{$W_3^q$};
            \PutText{7}{5}{1}{$W_4^q$};
            \PutText{7}{6}{1}{$W_5^q$};
            \PutText{7}{7}{1}{$W_6^q$};
            \PutText{7}{1}{1}{$\underset{\rightarrow}{\textrm{\smaller{user}}}$};
            \PutText{7}{1}{2}{\smaller{1}};
            \PutText{7}{1}{3}{\smaller{2}};
            \PutText{7}{1}{4}{\smaller{3}};
            \PutText{7}{1}{5}{\smaller{4}};
            \PutText{7}{1}{6}{\smaller{5}};
            \PutText{7}{1}{7}{\smaller{6}};
            \end{scope}
        \end{tikzpicture}
        \caption{Transmission vector $\hat{\Bx}(1)$}
        \label{fig:subhat1}
    \end{subfigure}
    \begin{subfigure}{.24\textwidth}
        \centering
        \begin{tikzpicture}[scale = 0.57]
            \begin{scope}<+->;
            \draw[step=1cm,thick,black!90] (1,0) grid (7,6);
            \draw[thin,black!70](0,0)to(1,0);
            \draw[thin,black!70](0,1)to(1,1);
            \draw[thin,black!70](0,2)to(1,2);
            \draw[thin,black!70](0,3)to(1,3);
            \draw[thin,black!70](0,4)to(1,4);
            \draw[thin,black!70](0,5)to(1,5);
            \draw[thin,black!70](0,6)to(1,6);
            \draw[thin,black!70](1,6)to(1,7);
            \draw[thin,black!70](2,6)to(2,7);
            \draw[thin,black!70](3,6)to(3,7);
            \draw[thin,black!70](4,6)to(4,7);
            \draw[thin,black!70](5,6)to(5,7);
            \draw[thin,black!70](6,6)to(6,7);
            \draw[thin,black!70](7,6)to(7,7);
            \end{scope}
            \begin{scope}
            
            \FillBlack{7}{2}{2};
            \FillBlack{7}{3}{3};
            \FillBlack{7}{4}{4};
            \FillBlack{7}{5}{5};
            \FillBlack{7}{6}{6};
            \FillBlack{7}{7}{7};
            
            \FillGray{7}{2}{4};
            \FillGray{7}{2}{5};
            \FillGray{7}{4}{2};
            
            \PutText{7}{2}{1}{$W_1^q$};
            \PutText{7}{3}{1}{$W_2^q$};
            \PutText{7}{4}{1}{$W_3^q$};
            \PutText{7}{5}{1}{$W_4^q$};
            \PutText{7}{6}{1}{$W_5^q$};
            \PutText{7}{7}{1}{$W_6^q$};
            \PutText{7}{1}{1}{$\underset{\rightarrow}{\textrm{\smaller{user}}}$};
            \PutText{7}{1}{2}{\smaller{1}};
            \PutText{7}{1}{3}{\smaller{2}};
            \PutText{7}{1}{4}{\smaller{3}};
            \PutText{7}{1}{5}{\smaller{4}};
            \PutText{7}{1}{6}{\smaller{5}};
            \PutText{7}{1}{7}{\smaller{6}};
            \end{scope}
        \end{tikzpicture}
        \caption{Transmission vector $\hat{\Bx}(2)$}
        \label{fig:subhat2}
    \end{subfigure}
    \caption{Graphical representation of transmission vectors for the virtual network}
    \label{fig:subhatnoxor}
\end{figure}

For simplicity, let us use $W(1) \equiv A$, $W(2) \equiv B$, ..., and $W(6) \equiv F$ (note that it is still possible for some users to request the same file). Using the coded caching scheme in~\cite{salehi2020lowcomplexity} for the virtual network, the first and second transmission vectors (corresponding to the transmissions in the first and second time intervals) are built as
\begin{equation}
\label{eq:trans_vector_exmp_virtual}
\begin{aligned}
    \hat{\Bx}(1) &= A_2^1 \Bw_{3} + B_1^1 \Bw_3 + C_1^1 \Bw_2 , \\
    \hat{\Bx}(2) &= A_3^1 \Bw_4 + C_1^2 \Bw_4 + D_1^1 \Bw_3 ,
\end{aligned}
\end{equation}
where the brackets for the zero-forcing sets are ignored for notational simplicity.
For better clarification, graphical representations for $\hat{\Bx}(1)$ and $\hat{\Bx}(2)$ are provided in Figure~\ref{fig:subhatnoxor}. These representations follow the same principles as in~\cite{salehi2020lowcomplexity}, i.e., 
%
a black cell means (every subpacket of) the respective data part is available in the cache memory, and a gray cell indicates that (a subpacket of) the respective part is sent during the transmission. 

Now, using the stretching mechanism in~\eqref{eq:stretch_main},  the stretched version of $\hat{\Bx}(1)$ can be written as
\begin{equation}
\begin{aligned}
    {\Bx}(1) =& A_2^{1,1} \Bw_{\Sfs_3^1,\Sfs_3^2,\Sfs_1^2} + A_2^{1,2} \Bw_{\Sfs_3^1,\Sfs_3^2,\Sfs_1^1} \\
    & + B_1^{1,1} \Bw_{\Sfs_3^1,\Sfs_3^2,\Sfs_2^2} + B_1^{1,2} \Bw_{\Sfs_3^1,\Sfs_3^2,\Sfs_2^1} \\
    & + C_1^{1,1} \Bw_{\Sfs_2^1,\Sfs_2^2,\Sfs_3^2} + C_1^{1,2} \Bw_{\Sfs_2^1,\Sfs_2^2,\Sfs_3^1} ,
\end{aligned}
\end{equation}
where the brackets for the zero-forcing sets are ignored for notational simplicity. Similarly, for $\hat{\Bx}(2)$ we can write
\begin{equation}
\begin{aligned}
    {\Bx}(2) =& A_3^{1,1} \Bw_{\Sfs_4^1,\Sfs_4^2,\Sfs_1^2} + A_3^{1,2} \Bw_{\Sfs_4^1,\Sfs_4^2,\Sfs_1^1} \\
    & + C_1^{2,1} \Bw_{\Sfs_4^1,\Sfs_4^2,\Sfs_3^2} + C_1^{2,2} \Bw_{\Sfs_4^1,\Sfs_4^2,\Sfs_3^1} \\
    & + D_1^{1,1} \Bw_{\Sfs_3^1,\Sfs_3^2,\Sfs_4^2} + D_1^{1,2} \Bw_{\Sfs_3^1,\Sfs_3^2,\Sfs_4^1} .
\end{aligned}
\end{equation}
Note that in this example, zero-forcing the interference of one data stream on three other streams is possible as the spatial multiplexing gain of $L=4$ is achievable at the transmitter. The graphical representations for these transmissions are provided in Figures~\ref{fig:extended} and~\ref{fig:extended_second}. As can be seen, they seem like the stretched versions of the base representations provided in Figure~\ref{fig:subhatnoxor}, and hence, we use the term \textit{stretching} to denote the proposed set of operations.

\begin{figure}[t]
    \centering
    \begin{subfigure}{.48\textwidth}
        \centering
        \begin{tikzpicture}[scale = 0.57]
            \begin{scope}<+->;
            \draw[thin,black!70](0,0)to(1,0);
            \draw[thin,black!70](0,1)to(1,1);
            \draw[thin,black!70](0,2)to(1,2);
            \draw[thin,black!70](0,3)to(1,3);
            \draw[thin,black!70](0,4)to(1,4);
            \draw[thin,black!70](0,5)to(1,5);
            \draw[thin,black!70](0,6)to(1,6);
            
            \draw[thick,black!90](1,0)to(13,0);
            \draw[thick,black!90](1,1)to(13,1);
            \draw[thick,black!90](1,2)to(13,2);
            \draw[thick,black!90](1,3)to(13,3);
            \draw[thick,black!90](1,4)to(13,4);
            \draw[thick,black!90](1,5)to(13,5);
            \draw[thick,black!90](1,6)to(13,6);
            
            \draw[thin,black!60](1,7)to(13,7);
            
            \draw[thick,black!90](1,0)to(1,6);
            \draw[thin,black!70](1,6)to(1,8);
            
            \draw[very thin,black!60](2,0)to(2,7);
            
            \draw[thick,black!90](3,0)to(3,6);
            \draw[thin,black!70](3,6)to(3,8);
            
            \draw[very thin,black!60](4,0)to(4,7);
            
            \draw[thick,black!90](5,0)to(5,6);
            \draw[thin,black!70](5,6)to(5,8);
            
            \draw[very thin,black!60](6,0)to(6,7);
            
            \draw[thick,black!90](7,0)to(7,6);
            \draw[thin,black!70](7,6)to(7,8);
            
            \draw[very thin,black!60](8,0)to(8,7);
            
            \draw[thick,black!90](9,0)to(9,6);
            \draw[thin,black!70](9,6)to(9,8);
            
            \draw[very thin,black!60](10,0)to(10,7);
            
            \draw[thick,black!90](11,0)to(11,6);
            \draw[thin,black!70](11,6)to(11,8);
            
            \draw[very thin,black!60](12,0)to(12,7);
            
            \draw[thick,black!90](13,0)to(13,6);
            \draw[thin,black!70](13,6)to(13,8);
            \end{scope}
            \begin{scope}
            
            
            
            
            \FillBlack{7}{2}{2};
            \FillBlack{7}{2}{3};
            \FillBlack{7}{3}{4};
            \FillBlack{7}{3}{5};
            \FillBlack{7}{4}{6};
            \FillBlack{7}{4}{7};
            \FillBlack{7}{5}{8};
            \FillBlack{7}{5}{9};
            \FillBlack{7}{6}{10};
            \FillBlack{7}{6}{11};
            \FillBlack{7}{7}{12};
            \FillBlack{7}{7}{13};
            
            \FillGray{7}{3}{2};
            \FillGray{7}{3}{3};
            \FillGray{7}{2}{4};
            \FillGray{7}{2}{5};
            \FillGray{7}{2}{6};
            \FillGray{7}{2}{7};
            
            \PutText{7}{2}{1}{$W_1^q$};
            \PutText{7}{3}{1}{$W_2^q$};
            \PutText{7}{4}{1}{$W_3^q$};
            \PutText{7}{5}{1}{$W_4^q$};
            \PutText{7}{6}{1}{$W_5^q$};
            \PutText{7}{7}{1}{$W_6^q$};
            \PutText{7}{0}{2.5}{\smaller{User 1}};
            \PutText{7}{0}{4.5}{\smaller{User 2}};
            \PutText{7}{0}{6.5}{\smaller{User 3}};
            \PutText{7}{0}{8.5}{\smaller{User 4}};
            \PutText{7}{0}{10.5}{\smaller{User 5}};
            \PutText{7}{0}{12.5}{\smaller{User 6}};
            \PutText{7}{1}{2}{$\Sfs_1^1$};
            \PutText{7}{1}{3}{$\Sfs_1^2$};
            \PutText{7}{1}{4}{$\Sfs_2^1$};
            \PutText{7}{1}{5}{$\Sfs_2^2$};
            \PutText{7}{1}{6}{$\Sfs_3^1$};
            \PutText{7}{1}{7}{$\Sfs_3^2$};
            \PutText{7}{1}{8}{$\Sfs_4^1$};
            \PutText{7}{1}{9}{$\Sfs_4^2$};
            \PutText{7}{1}{10}{$\Sfs_5^1$};
            \PutText{7}{1}{11}{$\Sfs_5^2$};
            \PutText{7}{1}{12}{$\Sfs_6^1$};
            \PutText{7}{1}{13}{$\Sfs_6^2$};
            \end{scope}
        \end{tikzpicture}
    \end{subfigure}
    \caption{Graphical illustration of ${\Bx}(1)$}
    \label{fig:extended}
\end{figure}

\begin{figure}[t]
    \centering
    \begin{subfigure}{.48\textwidth}
        \centering
        \begin{tikzpicture}[scale = 0.57]
            \begin{scope}<+->;
            \draw[thin,black!70](0,0)to(1,0);
            \draw[thin,black!70](0,1)to(1,1);
            \draw[thin,black!70](0,2)to(1,2);
            \draw[thin,black!70](0,3)to(1,3);
            \draw[thin,black!70](0,4)to(1,4);
            \draw[thin,black!70](0,5)to(1,5);
            \draw[thin,black!70](0,6)to(1,6);
            
            \draw[thick,black!90](1,0)to(13,0);
            \draw[thick,black!90](1,1)to(13,1);
            \draw[thick,black!90](1,2)to(13,2);
            \draw[thick,black!90](1,3)to(13,3);
            \draw[thick,black!90](1,4)to(13,4);
            \draw[thick,black!90](1,5)to(13,5);
            \draw[thick,black!90](1,6)to(13,6);
            
            \draw[thin,black!60](1,7)to(13,7);
            
            \draw[thick,black!90](1,0)to(1,6);
            \draw[thin,black!70](1,6)to(1,8);
            
            \draw[very thin,black!60](2,0)to(2,7);
            
            \draw[thick,black!90](3,0)to(3,6);
            \draw[thin,black!70](3,6)to(3,8);
            
            \draw[very thin,black!60](4,0)to(4,7);
            
            \draw[thick,black!90](5,0)to(5,6);
            \draw[thin,black!70](5,6)to(5,8);
            
            \draw[very thin,black!60](6,0)to(6,7);
            
            \draw[thick,black!90](7,0)to(7,6);
            \draw[thin,black!70](7,6)to(7,8);
            
            \draw[very thin,black!60](8,0)to(8,7);
            
            \draw[thick,black!90](9,0)to(9,6);
            \draw[thin,black!70](9,6)to(9,8);
            
            \draw[very thin,black!60](10,0)to(10,7);
            
            \draw[thick,black!90](11,0)to(11,6);
            \draw[thin,black!70](11,6)to(11,8);
            
            \draw[very thin,black!60](12,0)to(12,7);
            
            \draw[thick,black!90](13,0)to(13,6);
            \draw[thin,black!70](13,6)to(13,8);
            \end{scope}
            \begin{scope}
            
            
            
            
            \FillBlack{7}{2}{2};
            \FillBlack{7}{2}{3};
            \FillBlack{7}{3}{4};
            \FillBlack{7}{3}{5};
            \FillBlack{7}{4}{6};
            \FillBlack{7}{4}{7};
            \FillBlack{7}{5}{8};
            \FillBlack{7}{5}{9};
            \FillBlack{7}{6}{10};
            \FillBlack{7}{6}{11};
            \FillBlack{7}{7}{12};
            \FillBlack{7}{7}{13};
            
            \FillGray{7}{4}{2};
            \FillGray{7}{4}{3};
            \FillGray{7}{2}{6};
            \FillGray{7}{2}{7};
            \FillGray{7}{2}{8};
            \FillGray{7}{2}{9};
            
            \PutText{7}{2}{1}{$W_1^q$};
            \PutText{7}{3}{1}{$W_2^q$};
            \PutText{7}{4}{1}{$W_3^q$};
            \PutText{7}{5}{1}{$W_4^q$};
            \PutText{7}{6}{1}{$W_5^q$};
            \PutText{7}{7}{1}{$W_6^q$};
            \PutText{7}{0}{2.5}{\smaller{User 1}};
            \PutText{7}{0}{4.5}{\smaller{User 2}};
            \PutText{7}{0}{6.5}{\smaller{User 3}};
            \PutText{7}{0}{8.5}{\smaller{User 4}};
            \PutText{7}{0}{10.5}{\smaller{User 5}};
            \PutText{7}{0}{12.5}{\smaller{User 6}};
            \PutText{7}{1}{2}{$\Sfs_1^1$};
            \PutText{7}{1}{3}{$\Sfs_1^2$};
            \PutText{7}{1}{4}{$\Sfs_2^1$};
            \PutText{7}{1}{5}{$\Sfs_2^2$};
            \PutText{7}{1}{6}{$\Sfs_3^1$};
            \PutText{7}{1}{7}{$\Sfs_3^2$};
            \PutText{7}{1}{8}{$\Sfs_4^1$};
            \PutText{7}{1}{9}{$\Sfs_4^2$};
            \PutText{7}{1}{10}{$\Sfs_5^1$};
            \PutText{7}{1}{11}{$\Sfs_5^2$};
            \PutText{7}{1}{12}{$\Sfs_6^1$};
            \PutText{7}{1}{13}{$\Sfs_6^2$};
            \end{scope}
        \end{tikzpicture}
    \end{subfigure}
    \caption{Graphical illustration of ${\Bx}(2)$}
    \label{fig:extended_second}
\end{figure}

Now, let us review the decoding process for each stream of user one as ${\Bx}(1)$ is transmitted. The received signal at this user can be written in the vector form as 
\begin{equation}
    \begin{aligned}
        \By_1(1) &= \BH_1 {\Bx}(1) + \Bz_1(1) \\
        &= A_2^{1,1} \BH_{1} \Bw_{\Sfs_3^1,\Sfs_3^2,\Sfs_1^2} + A_2^{1,2} \BH_{1} \Bw_{\Sfs_3^1,\Sfs_3^2,\Sfs_1^1} \\
        &\qquad+ \BH_1 \BI_1(1) +  \Bz_1(1) ,
    \end{aligned}
\end{equation}
where $\BI_1(i)$ is the interference term at user one at the first time interval, defined as
\begin{equation}
    \begin{aligned}
        \BI_1(1) &= B_1^{1,1} \Bw_{\Sfs_3^1,\Sfs_3^2,\Sfs_2^2} + B_1^{1,2} \Bw_{\Sfs_3^1,\Sfs_3^2,\Sfs_2^1} \\
        & + C_1^{1,1} \Bw_{\Sfs_2^1,\Sfs_2^2,\Sfs_3^2} + C_1^{1,2} \Bw_{\Sfs_2^1,\Sfs_2^2,\Sfs_3^1} .
    \end{aligned}
\end{equation}
The next step is to multiply the received signal $\By_1(1)$ by receive beamformers $\Bu_{1,1}$ and $\Bu_{1,2}$, in order to extract $\Sfs_1^1$ and $\Sfs_1^2$, respectively. After multiplication by $\Bu_{1,1}$ we get
\begin{equation}
    \begin{aligned}
        y_1^1(1) &= \Bu_{1,1}^H \By_1(1) = \Bh_{1,1}^H \Bx(1) + z_{1,1}(1) \\
        &= A_2^{1,1} \Bh_{1,1}^H \Bw_{\Sfs_3^1,\Sfs_3^2,\Sfs_1^2} + \Bh_{1,1}^H \BI_1(1) +  z_{1,1}(1) \; ,
    \end{aligned}
\end{equation}
where $\Bh_{1,1}$ and $z_{1,1}(1)$ are the equivalent channel vector and \ac{AWGN} for the first stream of user one during this transmission as defined in Section~\ref{section:sys_model}, and we have used the zero-force beamforming property
\begin{equation}
    \Bh_{1,1}^H \Bw_{\Sfs_3^1,\Sfs_3^2,\Sfs_1^1} = 0 \; .
\end{equation}
However, user one has all the terms $B_1^{1,1}$, $B_1^{1,2}$, $C_1^{1,1}$, and $C_1^{1,2}$ in its cache memory. Hence, with the equivalent channel multiplier knowledge, it can build and remove the interference term $\Bh_{1,1}^H \BI_1(1)$ from $y_1^1(1)$, to decode the requested term $A_1^{1,1}$ interference-free. Similarly, multiplying by $\Bu_{1,2}$ results in
\begin{equation}
    \begin{aligned}
        y_1^2(1) &= \Bu_{1,2}^H \By_1(1) = \Bh_{1,2}^H \Bx(1) + z_{1,2}(1) \\
        &= A_2^{1,2} \Bh_{1,2}^H \Bw_{\Sfs_3^1,\Sfs_3^2,\Sfs_1^1} + \Bh_{1,2}^H \BI_1(1) +  z_{1,2}(1) \; .
    \end{aligned}
\end{equation}
Again, user one can use its cache contents and the acquired equivalent channel multipliers to build and remove $\Bh_{1,2}^H \BI_1(1)$ from $y_1^2(1)$, and decode $A_1^{1,2}$ interference-free. As a result, user one receives two parallel data streams from the transmission of $\Bx(1)$. Similarly, users two and three each receive two parallel data streams, and hence, a total number of six data streams are delivered to the users during the first time interval. The same procedure is repeated during every transmission, and the increased \ac{DoF} of $Gt+L = 6$ becomes possible. Note that here, the cache-aided interference cancellation (i.e., removing $\Bh_{1,1}^H \BI_1(1)$ and $\Bh_{1,2}^H \BI_1(1)$) occurs before decoding $y_1^1(1)$ and $y_1^2(1)$, and hence, we have a signal-level coded caching scheme. Moreover, as $p \in [6]$, $q \in [3]$, and $g \in [2]$, the total subpacketization of the caching scheme in this example network is $6 \times 3 \times 2 = 36$.

\subsection{Performance and Complexity Analysis}
It can be easily verified that using the proposed elevation mechanism, if the baseline \ac{MISO} scheme achieves \ac{DoF} of $t+L$, the resulting \ac{MIMO} scheme can deliver $Gt+L$ parallel streams to the users during each transmission. This is because for the original network with parameters $K$, $t$, $L$, and $G$, the virtual \ac{MISO} network has the same number of $K$ users and coded caching gain $t$, but the transmitter-side multiplexing gain is reduced to $\eta = \frac{L}{G}$. So, each transmission vector in the virtual network delivers data to $t+\eta$ users simultaneously. However, according to~\eqref{eq:stretch_main}, during the stretching process every data stream in a transmission vector of the virtual network is replaced by $G$ data stream for the real network. So, in every transmission vector for the original \ac{MIMO} network we deliver
\begin{equation}
    G(t+\eta) = Gt + L
\end{equation}
parallel streams, and hence, the \ac{DoF} is increased to $Gt+L$. As discussed in~\cite{parrinello2019fundamental}, this \ac{DoF} value is also optimal among any linear coded caching scheme with uncoded cache placement and single-shot data delivery.

The subpacketization value of the resulting \ac{MIMO} scheme depends on the subpacketization requirement of the baseline \ac{MISO} scheme. If the subpacketization of the baseline scheme is $f(K,t,L)$, where $K$, $t$, and $L$ represent the user count, the coded caching gain, and the transmitter-side multiplexing gain, respectively, the resulting \ac{MIMO} scheme from the proposed elevation mechanism will have the subpacketization of
\begin{equation}
    G \times f(K,t,\eta) \; .
\end{equation}
This is because the subpacketization requirement for the virtual network is $f(K,t,\eta)$, and during the elevation process, each subpacket is further split into $G$ smaller parts. So, for example, if the low-subpacketization scheme in~\cite{salehi2020lowcomplexity} is selected as the baseline \ac{MISO} scheme, the subpacketization requirement would be
\begin{equation}
    G \times K(t+\eta) = K(Gt+L) \; .
\end{equation}
However, the scheme in~\cite{salehi2020lowcomplexity} is applicable to the virtual network only if $\eta \ge t$, or $L \ge Gt$. On the other hand, if we select the original \ac{MISO} scheme in~\cite{shariatpanahi2018multi} as the baseline, the resulting \ac{MIMO} scheme would have the subpacketization requirement of
\begin{equation}
    G \times \binom{K}{t} \binom{K-t-1}{\eta-1} \; ,
\end{equation}
and the scheme would be applicable regardless of $t$, $L$, and $G$ values as long as $\eta$ is an integer. In summary, depending on the selection of the baseline scheme, the resulting \ac{MIMO} scheme can benefit from a very low subpacketization complexity. The subpacketization growth with the user count can even be linear, if the baseline scheme is chosen as the one in~\cite{salehi2020lowcomplexity}.

\section{Conclusion and Future Work}
In this paper, we investigated how coded caching schemes can be applied to multi-input multi-output (MIMO) communication setups. We introduced a novel mechanism to elevate any multi-input single-output (MISO) scheme in the literature to be applicable to MIMO settings. The elevation mechanism keeps the properties of the baseline scheme (e.g., the low subpacketization requirement) while achieving an increased coded caching gain. We also used results from the shared-cache model to claim the optimality of the achievable degrees of freedom (DoF) of the resulting scheme among the class of linear, single-shot caching schemes with uncoded data placement. Overall, the proposed elevation mechanism allows designing efficient, low-complexity coded caching schemes for any MIMO network.

The discussions in this paper were focused on communications in the high-SNR regime where the DoF metric is an appropriate performance indicator. However, a thorough analysis of the performance at the finite-SNR regime is due for later work. Also, investigating the applicability of the proposed scheme under dynamic network conditions where users can freely join/leave the network is part of the ongoing research.


\bibliographystyle{IEEEtran}
\bibliography{references}

\begin{thebibliography}{10}
\providecommand{\url}[1]{#1}
\csname url@samestyle\endcsname
\providecommand{\newblock}{\relax}
\providecommand{\bibinfo}[2]{#2}
\providecommand{\BIBentrySTDinterwordspacing}{\spaceskip=0pt\relax}
\providecommand{\BIBentryALTinterwordstretchfactor}{4}
\providecommand{\BIBentryALTinterwordspacing}{\spaceskip=\fontdimen2\font plus
\BIBentryALTinterwordstretchfactor\fontdimen3\font minus
  \fontdimen4\font\relax}
\providecommand{\BIBforeignlanguage}[2]{{%
\expandafter\ifx\csname l@#1\endcsname\relax
\typeout{** WARNING: IEEEtran.bst: No hyphenation pattern has been}%
\typeout{** loaded for the language `#1'. Using the pattern for}%
\typeout{** the default language instead.}%
\else
\language=\csname l@#1\endcsname
\fi
#2}}
\providecommand{\BIBdecl}{\relax}
\BIBdecl

\bibitem{cisco2018cisco}
V.~N.~I. Cisco, ``{Cisco visual networking index: Forecast and trends,
  2017--2022},'' \emph{White Paper}, vol.~1, 2018.

\bibitem{mahmoodi2021non}
H.~B. Mahmoodi, M.~Salehi, and A.~T{\"{o}}lli, ``{Non-Symmetric Coded Caching
  for Location-Dependent Content Delivery},'' \emph{arXiv preprint
  arXiv:2102.02518}, 2021.

\bibitem{maddah2014fundamental}
M.~A. Maddah-Ali and U.~Niesen, ``{Fundamental limits of caching},'' \emph{IEEE
  Transactions on Information Theory}, vol.~60, no.~5, pp. 2856--2867, 2014.

\bibitem{rajatheva2020white}
N.~Rajatheva, I.~Atzeni, E.~Bjornson, A.~Bourdoux, S.~Buzzi, J.-B. Dore,
  S.~Erkucuk, M.~Fuentes, K.~Guan, Y.~Hu, and {others}, ``{White Paper on
  Broadband Connectivity in 6G},'' \emph{arXiv preprint arXiv:2004.14247},
  2020.

\bibitem{shariatpanahi2016multi}
S.~P. Shariatpanahi, S.~A. Motahari, and B.~H. Khalaj, ``{Multi-server coded
  caching},'' \emph{IEEE Transactions on Information Theory}, vol.~62, no.~12,
  pp. 7253--7271, 2016.

\bibitem{shariatpanahi2018physical}
S.~P. Shariatpanahi, G.~Caire, and B.~Hossein~Khalaj, ``{Physical-Layer Schemes
  for Wireless Coded Caching},'' \emph{IEEE Transactions on Information
  Theory}, vol.~65, no.~5, pp. 2792--2807, 2019.

\bibitem{lampiris2018resolving}
\BIBentryALTinterwordspacing
E.~Lampiris and P.~Elia, ``{Resolving a feedback bottleneck of multi-antenna
  coded caching},'' \emph{arXiv}, 2018. [Online]. Available:
  \url{http://arxiv.org/abs/1811.03935}
\BIBentrySTDinterwordspacing

\bibitem{naderializadeh2017fundamental}
N.~Naderializadeh, M.~A. Maddah-Ali, and A.~S. Avestimehr, ``{Fundamental
  Limits of Cache-Aided Interference Management},'' \emph{IEEE Transactions on
  Information Theory}, vol.~63, no.~5, pp. 3092--3107, 2017.

\bibitem{shariatpanahi2018multi}
\BIBentryALTinterwordspacing
S.~P. Shariatpanahi and B.~H. Khalaj, ``{On Multi-Server Coded Caching in the
  Low Memory Regime},'' \emph{arXiv preprint arXiv:1803.07655}, pp. 1--12,
  2018. [Online]. Available: \url{https://arxiv.org/pdf/1803.07655.pdf}
\BIBentrySTDinterwordspacing

\bibitem{tolli2017multi}
A.~T{\"{o}}lli, S.~P. Shariatpanahi, J.~Kaleva, and B.~H. Khalaj,
  ``{Multi-antenna interference management for coded caching},'' \emph{IEEE
  Transactions on Wireless Communications}, vol.~19, no.~3, pp. 2091--2106,
  2020.

\bibitem{yan2018placement}
Q.~Yan, X.~Tang, Q.~Chen, and M.~Cheng, ``{Placement Delivery Array Design
  Through Strong Edge Coloring of Bipartite Graphs},'' \emph{IEEE
  Communications Letters}, vol.~22, no.~2, pp. 236--239, 2018.

\bibitem{yan2017placement}
Q.~Yan, M.~Cheng, X.~Tang, and Q.~Chen, ``{On the placement delivery array
  design for centralized coded caching scheme},'' \emph{IEEE Transactions on
  Information Theory}, vol.~63, no.~9, pp. 5821--5833, 2017.

\bibitem{shangguan2018centralized}
C.~Shangguan, Y.~Zhang, and G.~Ge, ``{Centralized coded caching schemes: A
  hypergraph theoretical approach},'' \emph{IEEE Transactions on Information
  Theory}, vol.~64, no.~8, pp. 5755--5766, 2018.

\bibitem{lampiris2018adding}
E.~Lampiris and P.~Elia, ``{Adding transmitters dramatically boosts
  coded-caching gains for finite file sizes},'' \emph{IEEE Journal on Selected
  Areas in Communications}, vol.~36, no.~6, pp. 1176--1188, 2018.

\bibitem{salehi2020lowcomplexity}
M.~Salehi, E.~Parrinello, S.~P. Shariatpanahi, P.~Elia, and A.~T{\"{o}}lli,
  ``{Low-Complexity High-Performance Cyclic Caching for Large MISO Systems},''
  \emph{arXiv preprint arXiv:2009.12231}, 2020.

\bibitem{salehi2019coded}
M.~Salehi, A.~T{\"{o}}lli, and S.~P. Shariatpanahi, ``{A Multi-Antenna Coded
  Caching Scheme with Linear Subpacketization},'' in \emph{2020 IEEE
  International Conference on Communications (ICC)}, 2019, pp. 1--6.

\bibitem{salehi2020diagonal}
M.~Salehi and A.~Tolli, ``{Diagonal Multi-Antenna Coded Caching for Reduced
  Subpacketization},'' in \emph{2020 IEEE Global Communications Conference
  (GLOBECOM)}.\hskip 1em plus 0.5em minus 0.4em\relax IEEE, 2020, pp. 1--6.

\bibitem{mohajer2020miso}
S.~Mohajer and I.~Bergel, ``{MISO Cache-Aided Communication with Reduced
  Subpacketization},'' in \emph{ICC 2020-2020 IEEE International Conference on
  Communications (ICC)}.\hskip 1em plus 0.5em minus 0.4em\relax IEEE, 2020, pp.
  1--6.

\bibitem{salehi2019subpacketization}
M.~Salehi, A.~Tolli, S.~P. Shariatpanahi, and J.~Kaleva,
  ``{Subpacketization-rate trade-off in multi-antenna coded caching},'' in
  \emph{2019 IEEE Global Communications Conference, GLOBECOM 2019 -
  Proceedings}.\hskip 1em plus 0.5em minus 0.4em\relax IEEE, 2019, pp. 1--6.

\bibitem{salehi2020coded}
M.~Salehi, A.~Tolli, and S.~P. Shariatpanahi, ``{Coded Caching with Uneven
  Channels: A Quality of Experience Approach},'' in \emph{2020 IEEE
  International Workshop on Signal Processing Advances in Wireless
  Communications (SPAWC)}, 2020, pp. 1--5.

\bibitem{parrinello2019fundamental}
E.~Parrinello, A.~{\"{U}}nsal, and P.~Elia, ``{Fundamental Limits of Coded
  Caching with Multiple Antennas, Shared Caches and Uncoded Prefetching},''
  \emph{IEEE Transactions on Information Theory}, 2019.

\bibitem{parrinello2020extending}
E.~Parrinello, P.~Elia, and E.~Lampiris, ``{Extending the Optimality Range of
  Multi-Antenna Coded Caching with Shared Caches},'' in \emph{2020 IEEE
  International Symposium on Information Theory (ISIT)}.\hskip 1em plus 0.5em
  minus 0.4em\relax IEEE, 2020, pp. 1675--1680.

\end{thebibliography}

\end{document}